\documentstyle[prl,aps,epsf] {revtex}
\begin{document}
\title{Testing quantum correlations in a confined
atomic cloud by scattering fast atoms}
\author{A.B. Kuklov$^1$ and B.V. Svistunov$^2$}
\address{$^1$ Department of Engineering Science and Physics,
The College of  Staten Island, CUNY,
     Staten Island, NY 10314}
\address{$^2$ Russian Research Center ``Kurchatov Institute",
123182 Moscow, Russia}

\date{\today}
\maketitle

\begin{abstract}
We suggest measuring  one-particle density matrix
of a trapped ultracold atomic cloud by scattering
fast atoms in a pure momentum state off the cloud.  
The lowest-order probability of the inelastic process,
resulting in a pair of outcoming fast
atoms for each incoming one, turns out to be given by 
a Fourier transform of the density matrix. Accordingly,
important information about
quantum correlations can be deduced directly from
the differential scattering cross-section. 
A possible design of the atomic detector is also discussed.
\\

\end{abstract}
\vskip0.5 cm

Successful advances in achieving collective quantum states 
in confined clouds of alkaline atoms \cite{BEC} 
and in atomic hydrogen \cite{HYDRO} make possible studying
quantum coherent properties of these systems, as well as the
revealing fundamental kinetic processes leading to the 
formation of the coherence.
In Ref.\cite{GROW}, a dynamics of the condensate growth has
been observed by means of detecting a density increase 
usually associated with the condensate formation in the trap.
The coherence of the condensate has been demonstrated  
directly by letting two released condensates form
the interference  pattern \cite{COHER}. 

Mechanism of formation of quantum correlations is 
a matter of great attention and controversy. An emergence of
the condensate due to either lowering temperature or reaching
an equilibrium after fast quenching is associated with 
a formation of the off-diagonal long-range order
(ODLRO) \cite{ODLRO}. A primary object
displaying such an order is the one-particle density matrix
(OPDM) $\rho ({\bf x}_1,{\bf x}_2)$. Typical 
distances over which these correlations become important
are comparable with the interatomic spacing $r_a$. Consequently,
an ``early" detection of such emerging correlations is very difficult
to achieve by light with the wavelength 
$\lambda \gg r_a$ usually employed for probing the cloud density. 
 In Ref.\cite{WALLS}, it has been suggested that a
resonant fluorescence of an external atom identical to those
forming the condensate is sensitive to the relative phase of two
condensates. However, this method is not suitable for testing 
distances shorter than $\lambda $.
Information about short-range density correlations 
(at distances $ < r_a$) can, in principle, be obtained
from the absorption of detuned resonant light \cite{ABSORB}.
The change in local $m$-body density correlations 
(the so-called $m!$-effect \cite{M!}) can be seen by measuring 
recombination rates, and this has been already done experimentally 
for the equilibrium case \cite {Burt}. However, measurements of
the correlation length $r_c$ of the forming ODLRO seem very
unlikely to be achieved by these methods. 

Thus it is tempting to have a tool which could make possible seeing
the OPDM directly without limitations on the accessible distances. 

Currently, great efforts are being dedicated to a creation
of a controllable source of coherent atoms -- atomic
LASER (see Ref.\cite{LASER} and references therein). 
In Ref.\cite{CANNON}, a mechanism of accelerating neutral
atoms has been proposed. Hence, it is very likely that a source of fast 
and coherent atoms will be available soon.
In this paper, we suggest a method of detecting the OPDM which relies 
on inelastic scattering of such atomic beam off the atomic cloud.  

We note that methods of scattering of neutrons \cite{NEUT1,NEUT2}
and He atoms \cite{ATOMS} off liquid He are well known. In Ref.
\cite{NEUT1}, the Impulse Approximation has been suggested to
employ for interpreting the differential cross-section of
fast neutrons. In such an approximation, it is possible to
relate the momentum transfer distribution to some
integral of the population factor in He. However,  
liquid He is a strongly interacting system where the
gas parameter $\xi =na^3$ ($n$ is the density and $a$ is 
the scattering length) is not small. Accordingly,
multiparticle excitations dominate in the
final-state channel, which makes a direct
measurement of the OPDM impossible. Certain
assumptions about the role of the final-state effects
should be made \cite{NEUT2}. 
In contrast to this, the gas parameter in the trapped
atomic condensates can be as small as $\xi \sim 10^{-5}$.
Accordingly, the mean free path is 
~$\approx a/\xi \sim 10^{-2}$ cm, which 
greatly exceeds the mean particle separation, and can become
greater then the cloud size. 
This implies that a contribution of the multiparticle
excitations can be safely neglected as long as a wavelength
of the incoming atom is much smaller than the interparticle
separation in the atomic cloud. Under these conditions,
as we will discuss below,
it becomes possible to measure the OPDM directly.
 
If the mass of the incoming fast atom is comparable
to that of the atoms forming a cloud (or, in particular,  
the fast atom is just identical to the atoms of the cloud), 
the lowest-order inelastic scattering event is a production
of two fast outcoming atoms (the size of atomic cloud is supposed
to be small enough to neglect multiple scattering). 
The quantum-mechanical probability
of this process turns out to be proportional 
to a Fourier transform of the OPDM. 
Another process - an elastic scattering of
a single fast atom off the cloud - also exists. 
In this case one has a single fast atom in the 
final state, in a complete analogy with elastic scattering of light.
In principle, the elastic processes might mask the inelastic ones. 
However, if the momentum of incoming atom, $k$, is much larger 
than the typical momentum of the the cloud, $\sim 1/r_c$, then
the elastic-scattering angles are small in the parameter 
$(k r_c)^{-1} \ll 1$. This feature allows one to distinguish between the 
two processes. One more feature, arising at $kr_c \gg 1$ from
the conservation laws, is the fact that the angle between
the two created fast atoms is always close to $\pi /2$. This
facilitates identification of pairs resulted from one and the
same scattering event.

Apart from extremely small uncertainty of the momentum of 
the incident atom, the method we suggest implies 
that the total momentum of the outcoming pair can be 
detected with a sufficient precision. 
Thus, a reliable atomic detector is required. 
Therefore, we will also discuss a possible design of
such a detector. 

Now let us derive an expression for the inelastic cross-section 
in the lowest order with respect to the two-body interaction.
For the sake of simplicity, we will not consider the 
complexity of all possible scattering channels and
will concentrate on 
a simplest case of spin-polarized bosons, 
when the incoming fast atom is identical to 
particles forming the cloud, and its spin 
polarization is the same as that of the cloud. 
We emphasize that the validity of the suggested method
relies on a possibility to have the incident momentum
$k$ obeying the relation

\begin{equation}
\xi^{1/3} \ll ka \ll 1 \; .
\label{eq:ka}
\end{equation}
\noindent 
Then, one can represent the interaction Hamiltonian in a 
traditional form ($\hbar=1$)

\begin{equation}
H_{int} \, = \, {u_0 \over 2}\int d{\bf x} \,
\Psi^{\dagger}({\bf x})\Psi^{\dagger}({\bf x})
\Psi ({\bf x})\Psi ({\bf x}) \; ,  \quad u_0={4\pi a\over m} \; ,
\label{eq:V_simple}
\end{equation}
\noindent
with $m$ standing for the atomic mass. The total field $\Psi $
can be subdivided into the low- and the high-energy parts, 
$\psi$ and $\psi '$, respectively:

\begin{equation}
\Psi = \psi + \psi' \; , \quad \psi' =\sum_{\bf k} a_{\bf k} \,
{\rm e}^{i{\bf kx}} \; ,
\label{eq:Psi}
\end{equation}
where the incident and the scattering states
are described in terms of plane waves 
normalized to unit volume; $a_{\bf k}$ destroys the high-energy particle
with the momentum ${\bf k}$. A substitution
of Eq.~(\ref{eq:Psi}) into Eq.~(\ref{eq:V_simple})
and selection of the terms that describe
a process involving one incident fast atom
carrying momentum ${\bf k}$ 
and two ejected fast atoms carrying momenta
${\bf k}_1$ and ${\bf k}_2$,
as well as its reverse, yield

\begin{equation}
H'_{int} \, = \, u_0\sum_{{\bf k}_1,{\bf k}_2,{\bf k}}
a^{\dagger}_{{\bf k}_1} a^{\dagger}_{{\bf k}_2} a_{\bf k}
\int d{\bf x} \, {\rm e}^{-i{\bf qx}}\psi \, + \, \mbox{H.c.} \; ,
\label{eq:H'}
\end{equation}
\noindent
where ~${\bf q}={\bf k}_1 + {\bf k}_2 - {\bf k}$~ is the 
transferred momentum.
The double-differential cross-section for scattering with 
given energy and momentum transfers, $\omega $ and ${\bf q}$,
in the lowest order with respect to $H'_{int}$ 
is given by the Golden rule formula

\begin{equation}
\displaystyle W({\bf q},\omega)=
{2 m u_0^2 \over k}\int {d{\bf k}_1\over (2\pi )^3} \,
\delta (\omega - \omega_{fi})
\int \! \! \int d{\bf x}_1 d{\bf x}_2\int_{-\infty}^{\infty} dt \,
{\rm e}^{i{\bf q}({\bf x}_1-{\bf x}_2)-i\omega t} \,
\rho ({\bf x}_1, t; {\bf x}_2, 0) \; , 
\label{eq:W_general}
\end{equation}
\noindent
where $\rho ({\bf x}_1, t_1; {\bf x}_2, t_2)= 
\langle \psi^{\dagger}({\bf x}_1, t_1)
\psi ({\bf x}_2, t_2)\rangle $ 
[Note that unlike the uniformity in time, 
~$\rho ({\bf x}_1, t_1; {\bf x}_2, t_2)=
\rho ({\bf x}_1, t_1-t_2; {\bf x}_2, 0)$,
the space uniformity cannot be, in general, assumed as long 
as a trapping potential exists.]; $\omega_{fi}$
is the difference of the kinetic energies 
of the fast atoms in the final and initial states:

\begin{equation}
\omega_{fi} = {1 \over m} \left[ {q^2\over 2}+ {\bf qk}+
 k^2_1- ({\bf q} + {\bf k}){\bf k_1} \right] \; ,
\label{eq: omega}
\end{equation}
\noindent
with ${\bf k}_1$ being the momentum of one of the outcoming
atoms while the momentum of the second one, ${\bf k}_2$,
is fixed by the relation 
${\bf k}_2= {\bf q}+{\bf k} - {\bf k}_1$. 

The integration over ${\bf k}_1$ in Eq.~(\ref{eq:W_general})
can be carried out explicitly. However, first
we notice that the requirement of large $k$ means that the values of 
$q$ and $\omega$, which are effectively selected by the correlator 
$\rho ({\bf x}_1, t; {\bf x}_2, 0)$ in the right-hand side of 
Eq.~(\ref{eq:W_general}), satisfy the conditions $q \ll k$ and
$|\omega | \ll  k^2 /m$. This immediately leads to the approximation
$\delta (\omega - \omega_{fi}) \approx m \, \delta (k_1^2 - {\bf k k}_1 )$,
which yields
 
\begin{equation}
\displaystyle W({\bf q},\omega)=
4a^2 \int \! \! \int d{\bf x}_1 d{\bf x}_2\int_{-\infty}^{\infty} dt \,
{\rm e}^{i{\bf q}({\bf x}_1-{\bf x}_2)-i\omega t} \, 
\rho ({\bf x}_1, t; {\bf x}_2, 0) \; . 
\label{eq:W}
\end{equation}
\noindent
We thus see that the double-differential cross-section $W({\bf q},\omega)$
is almost directly related to the dynamic correlator
$\rho ({\bf x}_1, t; {\bf x}_2, 0)$ which, as is known, contains
rather rich information about the system, including, for one thing,
 the elementary
excitation spectrum. Confining ourselves to the static correlations,
described by OPDM $\rho ({\bf x}_1,{\bf x}_2) = 
\rho ({\bf x}_1, 0; {\bf x}_2, 0)$, we arrive at even more simple relation
\begin{equation}
\displaystyle W({\bf q})=
8 \pi a^2 \int \! \! \int d{\bf x}_1 d{\bf x}_2 \, 
{\rm e}^{i{\bf q}({\bf x}_1-{\bf x}_2)} \rho ({\bf x}_1,{\bf x}_2) 
\label{eq:W_general_2}
\end{equation}
\noindent
in terms of the differential cross-section 
$W({\bf q})=\int d\omega \, W({\bf q},\omega)$.

Consider the structure of $W({\bf q})$ in the most characteristic cases. 
In the case of {\it pure Bose-Einstein condensate}, when
$\rho ({\bf x}_1,{\bf x}_2) = \Phi^*({\bf x}_1) \Phi ({\bf x}_2)$
[$\Phi ({\bf x})$ is the condensate wave-function], 
\begin{equation}
\displaystyle  W({\bf q}) \, = \,  8\pi a^2 | \Phi_{\bf q} |^2 \; , \; \; \;  
\Phi_{\bf q} = \int  d{\bf x} \, {\rm e}^{-i{\bf q}{\bf x}} \, 
\Phi ({\bf x}) \; . 
\label{eq:W_c}
\end{equation}
\noindent
In particular, if $\Phi_{\bf q}$ is real, Eqs.~(\ref{eq:W_c}) can
be reversed, and correspondingly $\Phi ({\bf x})$ can be restored from 
$W({\bf q})$. 

Another instructive case is the {\it axially symmetric quantum 
vortex}. A presence of a single vortex in a center of 
the axially symmetric condensate drastically changes
the scattering pattern. Indeed, in this situation, 
$\Phi =\exp(i\theta )\sqrt{n(r,z)}$, where $\theta $
is the axial angle and $n(r,z)$ stands for the axially 
symmetric condensate density as a function
of the distances $r,\, z$ perpendicular to the axis
and along the axis, respectively. Accordingly,
Eq.~(\ref{eq:W_c}) shows
that $W({\bf q})=0$ for ${\bf q}$ directed along the 
vortex line. The differential cross-section becomes
finite as long as there is  a component 
${\bf q}_{\perp}$ of ${\bf q}$ perpendicular
to the axis or the vortex displaces from the condensate
center. In the first case, $W({\bf q})\sim q_{\perp}^2$
for $ q_{\perp}\to 0$.

Quite similar to the quantum vortex is the case of
the {\it supercurrent state} of a toroidal Bose condensate.
The suppression of $W({\bf q})$ for ${\bf q}$ perpendicular
to the plane of the torus allows distinguishing the
supercurrent state from the currentless genuine ground state.

In the absence of condensate, or for the above-the-condensate
part of OPDM, one normally deals with the so-called 
{\it quasi-homogeneous regime}, when typical inverse momentum
of particles is much less then the scale of density variation.
In this case, it is reasonable to introduce the variables
~${\bf r} = {\bf x}_2 - {\bf x}_1$~ and 
~${\bf R} = ({\bf x}_1 + {\bf x}_2)/2$, and to represent
$W({\bf q})$ as 
\begin{equation}
\displaystyle  W({\bf q})= 8\pi a^2 \int d {\bf R} \, 
\rho_{\bf q}({\bf R})  \; , \; \; \;  
\rho_{\bf q}({\bf R}) = \int  d{\bf x} \,  {\rm e}^{-i{\bf q}{\bf x}} \,
\rho({\bf x},{\bf R}) \; , 
\label{eq:W_qh}
\end{equation}
\noindent
since in the quasi-homogeneous regime, OPDM in the Wigner representation,
$\rho_{\bf q}({\bf R})$, has a semiclassical meaning of local
(at the point ${\bf R}$) distribution of the particle momentum ${\bf q}$.
We thus see that in this case $W({\bf q})$ yields spatially
averaged momentum distribution. Moreover, without contradiction
with the uncertainty principle, this averaging can be 
partially (totally) removed by collimating the incident (both
the incident and outcoming) beams. Even without removing
the averaging, $W({\bf q})$ contains valuable information about
long-range correlations in the system, since the the averaging 
does not affect the order-of-magnitude value of the correlation
radius $r_c$ (equal to the inverse typical momentum $k_c$).
If the system contains both condensate and quasi-homogeneous
above-the-condensate fraction, the cross-section $W({\bf q})$
is given by a combination of Eqs.~(\ref{eq:W_c}) and (\ref{eq:W_qh}).

Now let us discuss one possible method of detecting
a total momentum of an outcoming pair of atoms.
There is a significant
requirement for such a detector: an uncertainty of
detecting the momentum transfer must be less than
the inverse correlation length $r_c$ in the cloud.
This imposes a limitation on the detector size $R_D$.
Indeed, let us suppose that cells sensitive to 
the arrival of the scattered atoms are located
on a sphere of the radius $R_D$, and the atomic
cloud of the size $R \ll R_D$ is at the center of
this sphere. Then, the uncertainty in the 
scattering angle is $R/R_D$. This produces
the uncertainty in the momentum ~$\sim kR/R_D$. This uncertainty must be
much less than $1/r_c$, or

\begin{equation}
R_D\, \gg \, kr_c \, R \; .
\label{eq:R_D}
\end{equation}
\noindent
It can be shown that the absolute values of the 
scattered momenta measured
by the time of flight are also subjected to the
same uncertainty. Thus, the condition (\ref{eq:R_D})
determines a precision of the time-of-flight measurements as well.

Our method implies that two outcoming atoms produced
by one incoming atom can be identified. As mentioned
above, this identification can be done due to
the $\approx \pi /2$ angle between the scattered
atoms. However, if the incoming
coherent beam produces too many scattered pairs,
the erroneous identification is very likely. Let us
derive a limitation on the number $N_s$ of the scattered
pairs per typical time of flight,
which would keep the momentum uncertainty
less than $1/r_c$. The scattering events
under consideration should occur with approximately
equal probability at any angle inside
the $4\pi $ solid angle, so that the angular
area occupied by a single event is
$4\pi /N_s$. It should be 
much less than $2\pi/r_ck$ which determines the
angular area of the strip where the second
atom of a pair can be found, the direction of
the first being fixed. Thus, 

\begin{equation}
N_s \ll 2 k r_c \; .
\label{eq:N_s}
\end{equation}
\noindent
If a typical speed of the fast atoms is 10cm/s, and
$r_c\approx 10^{-3}$ cm, $R\approx 10^{-2}$ cm,
the conditions (\ref{eq:R_D}) and (\ref{eq:N_s}) 
yield $R_D \gg 10$ cm and $N_s \ll 10^3$.
Accordingly, a typical time of flight
will be longer than 1 s. 

Now let us discuss a mechanism 
of detection of a single neutral atom.
We suggest employing resonant atomic  fluorescence in the
evanescent field of light propagating inside a 
waveguide. Specifically, light sensitive cells
are mounted on a side
of a long waveguide
so that light propagating inside 
the waveguide does not excite
these  cells. If holes are made through the cells
and the waveguide, the light will remain confined
as long as a diameter of the hole is much smaller
than the wavelength of light.
However, a single atom may enter the hole and
feel the resonant field inside the hole, or in the close
vicinity of the entrance to it, due to the evanescent
field of the light. Consequently, 
the atom will reemit resonantly one or several photons.
 These
photons can then be detected by the nearest cell indicating
an arrival of an atom at a specific location
at the detector surface. Accordingly,
a scattering angle  
of the coming atom can be identified. 
The velocity of the atom
can be deduced from the time of flight. 

A probability $P_D$ for atom to penetrate
into a hole
 is simply the area  
$A_h$ of the holes per unit area of the detector
exposed to the atomic flux. In order to detect
an atom in the hole, the atom must reemit at least
few photon which are captured by the cell mounted
outside the waveguide.
Before we estimate
a number $N_D$ of the reemitted photons which 
can be detected, 
we note that the photons remitted by the atom, 
which has penetrated deeply inside
the waveguide, should remain confined inside it, and
therefore they will not be detected by the cell. Only
those photons, which are reemitted by the coming atom
while being close to the hole entrance, will be scattered
almost isotropically and can be absorbed by the cell. 
Taking into account that 
the penetration length of the evanescent
field is comparable with the wavelength $\lambda $
 of light, one can find the time $t_D\approx \lambda /v$
during which the atom moving at speed $v$ 
is subjected to the evanescent field and reemits
light isotropically.
Then, the number of photons reemitted during this
time is $N_{D}\approx \gamma t_D= \gamma 
\lambda /v$,
where $\gamma$ stands for the natural width
of the line. Choosing
typical values $\lambda = 700$nm, $\gamma \approx 10^7$ Hz
and $v=10$ cm/s one finds $N_{ph}= 70$. Some
geometrical factor of the order of one should reduce 
the number of photons
reaching the cell.  
Quantum efficiency
of photodetectors can be easily achieved to be
10\%-20\%. This implies that
once an atom entered a hole, it will be detected
with high certainty. Therefore, a probability
to detect an atom at a given position is just
$P_D$. In other words, if a diameter of each hole is
100 nm and two closest holes are 300 nm apart, 
the probability is $P_D \approx 0.1$. Thus, in order
to achieve a resolution
of at least .1 of the
typical momentum region in the scattering intensity,
one needs at least 10 scattering events
for each component of the momentum. 
This corresponds to 10 atoms
ejected from the cloud (plus 10 fast incident atoms).
Accordingly, for the 3D geometry, it translates
into $10^3/P_D\approx 10^4$ atoms in total. In a typical
condensate of $10^6$ atoms \cite{BEC},
such a bombardment by fast atoms will result in a
 depletion of the cloud by only 1\%. Evidently, this
method is not appropriate for clouds containing 
less than $10^5$ atoms. In the effective 2D
or 1D geometries, the same resolution 
requires much less numbers of the scattering events.
 Specifically, $10^3$ and $10^2$ of them for the 2D
and 1D geometries, respectively, 
will satisfy the above criteria of resolution. 

In conclusion, we have suggested a method of scattering
of fast atoms in a pure enough momentum state 
off a trapped atomic cloud in order to
test directly one-particle density matrix of this
cloud. The differential cross-section
of the inelastic process, when one incoming fast
atom produces two fast ones, allows measuring
the correlation length of the local off-diagonal order. 
This gives, in particular, a powerful tool for testing 
different scenarios of formation of the off-diagonal 
long-range order in the traps. 
This method can also be employed for detecting quantum
vortices and supercurrent states, 
as well as the effect of quantum depletion
of the condensate. The main principles and a design
of the detector of scattered atoms are suggested.


\begin{references}
\bibitem{BEC} 
 M.H. Anderson, J.R. Ensher, M.R. Matthews, C.E. Wieman,
and E.A. Cornell, Science {\bf 269}, 198 (1995);
C.C. Bradley, C.A. Sackett, J.J. Tollett, and R.G. Hulet,
 Phys. Rev. Lett.{\bf 75},      1687 (1995);
 K.B. Davis, M.-O. Mewes, M.R. Andrews, N.J. van Druten,
D.S. Durfee,
D.M. Kurn, and W. Ketterle, Phys. Rev. Lett. {\bf 75}, 3969 (1995).
\bibitem{HYDRO} 
B.G. Levi, Physics Today, October 1998,p.17;
A.I.Safonov, S.A. Vasilyev, I.S. Yasnikov, I.I. 
Lukashevich, and S. Jaakkola, 
Phys. Rev. Lett. {\bf 81}, 4545 (1998).
\bibitem{GROW}
H.-J. Miesner, D. M. Stamper-Kurn, M.R. Andrews,
D.S. Durfee, S. Inouye,  W. Ketterle,
Science {\bf 279}, 1005 (1998).
\bibitem{COHER}
M.R. Andrews, C.G. Townsend, H.-J. Miesner,
D.S. Durfee, D.M. Kurn, and W. Ketterle,
Science {\bf 275}, 637 (1997). 
\bibitem{ODLRO}
C.N. Yang,  Rev.Mod.Phys. {\bf 34}, 694 (1962);
W. Kohn, D. Sherrington, Rev.Mod.Phys. {\bf 42}, 1 (1970).
\bibitem{WALLS}
J. Ruostekoski, D.F. Walls, Phys.Rev. A {\bf 56}, 2996 (1997).
\bibitem{ABSORB}
Yu. Kagan, B.V. Svistunov, and G.V. Shlyapnikov,
JETP Lett., {\bf 48}, 56 (1988).
\bibitem{M!}
Yu. Kagan, B.V. Svistunov, and G.V. Shlyapnikov,
JETP Lett. {\bf 42}, 209 (1985). 
\bibitem{Burt} E.A. Burt, R.W. Christ, C.J. Myatt, M.J. Holland,
E.A. Cornell, and C.E. Wieman,
Phys. Rev. Lett. {\bf 79}, 337 (1997).
\bibitem{LASER}
H. Steck, M. Naraschewski, and H. Wallis,
Phys.Rev.Lett. {\bf 80}, 1 (1998).
\bibitem{CANNON}
V.I. Yukalov and E.P. Yukalova,
preprint cond-mat/9809103.
\bibitem{NEUT1}
P.C. Hohenberg and P.M. Platzman, Phys. Rev. {\bf 152},
198 (1966). 
\bibitem{NEUT2}
P.E. Sokol, in {\it Excitations in Two-Dimensional
and Three-Dimensional Quantum Fluids}, ed. A.F.G. Wyatt
and H.J. Lauter, Plenum Press, 1990, p. 47.
\bibitem{ATOMS}
D. O. Edwards, P.P. Fatorous et al., Phys. Rev. Lett.
{\bf 34}, 1153 (1975); V. U. Nayak, D. O. Edwards,
and N. Masuhara, Phys. Rev. Lett. {\bf 50}, 990 (1983);
A. K. Setty, J. W. Halley, and C.E. Campbell,
Phys. Rev. Lett. {\bf 79}, 3930 (1997);

\end{references}
\end{document}